\documentclass[prb,showpacs,floatfix,twocolumn]{revtex4}
\usepackage{graphicx}
\usepackage{amssymb}
\usepackage{dcolumn}
\usepackage{bm}
\begin{document}

\title{Finite width of quasi-static shear bands}

\author{E. A. Jagla}

\affiliation{Centro At\'omico Bariloche, Comisi\'on Nacional de Energ\'{\i}a At\'omica, 
(8400) Bariloche, Argentina}

\begin{abstract}

I study the average deformation rate of an amorphous material submitted to an external uniform shear strain rate, in the geometry known as the split-bottom configuration. The material is described using a stochastic model of plasticity
at a mesoscopic scale.
A shear band is observed to start at the split point at the bottom, and widen progressively towards the surface.
In a two-dimensional geometry the average statistical properties of the shear band look similar to those of the directed polymer model. In particular, the surface width of the shear band is found to scale with the system height $H$ as 
$H^\alpha $ with $\alpha=0.68\pm 0.02$. In more realistic three dimensional simulations the exponent changes to 
$\alpha=0.60\pm 0.02$  and the bulk profile of the width of the shear band is closer to a quarter of circle, as it was
observed to be the case in recent simulations of granular materials. 

\end{abstract}
\maketitle

\section{Introduction}

Amorphous and granular materials have a strong tendency to develop localized deformation when submitted to shear deformation. 
This deformation concentrates usually in narrow regions, called shear bands. Upon a
uniform rate of shear deformation, the material experiences a finite, stationary deformation rate within the 
shear band, whereas the deformation rate is vanishingly small outside it.\cite{libro1,libro2}
In some cases the existence of strain localization can be associated to a decreasing form of the stress as a function of strain rate in the material. This decreasing leads to instabilities that drive the formation of shear bands.

An interesting limiting case corresponds to the so called quasi-static shear banding. In this case the 
applied strain rate is vanishingly small, and the velocity field across the sample must change by a factor
when there is a change of the driving force by the same factor.
It seems that in this case, an stationary, finite width shear band can only be obtained through the
application of appropriate conditions at the boundaries that break the uniformity of the system. In fact, simulations
of homogeneous amorphous materials well below the glass temperature show the existence of shear band which however
thicken in time, eventually taking over the whole system. \cite{falk}

A particular example in which the shear band is induced through the use of appropriate boundary conditions is the split-bottom container geometry introduced in 
\cite{fenis1,fenis2} in the form of a modified Couette cell. In this configuration the bottom of the 
cell is split into two disks that rotate relatively to each other, and are attached to the inner and outer cylindrical wall of the cell. The sharp velocity interface at the bottom of the container 
induces an abrupt change of the velocity profile of the material close to this split line. 
Experiments using granular materials show that in these conditions the material inside the cell reaches (on average) a stable configuration in which the transition between particles moving rigidly with one or the other part of the bottom surface
occurs in a transition zone that defines the shear band. The form of this zone is well defined and non-singular even in the limit of `quasi-static' shear bands, i.e, when the relative velocity of both parts of the container is vanishingly small. 
However, when horizontal planes progressively away of the bottom are considered (and thus the influence of the split-bottom decreases), the width of the shear zone becomes larger, and no upper limit seems to exist.
The width of the shear band starts as zero right at the bottom of the container, and progressively
increases when moving upwards, being maximum at the free surface. The detailed study of the form of this profile is the subject of this paper. 

\begin{figure}[h]
\includegraphics[width=8cm,clip=true]{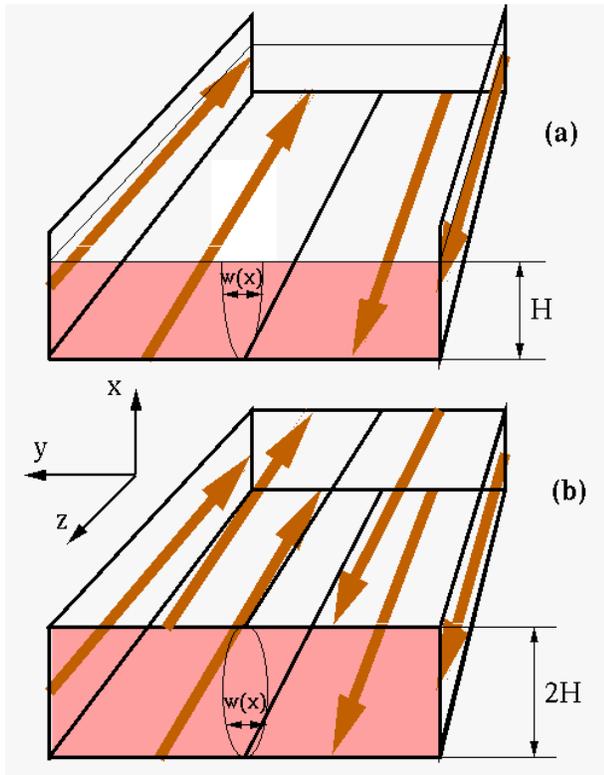}
\caption{(a) The `split bottom' geometry.
Left and right parts of the system are driven at a
very small velocity $v$ and $-v$, respectively. The top surface is free of force. The velocity of the 
material match these values close to the bottom surface, and interpolates continuously
in the bulk, as indicated schematically. (b) A `split top-bottom' geometry. Now the top surface is also split into two, like the bottom. Note the definition of the height $H$ in both cases, which is made in this way to allow a better comparison.
Even in the absence of gravity effects is not obvious that the configuration in (a) should be equivalent 
to half of the configuration in (b), as it is explained in the text.
}
\label{f1}
\end{figure}

The geometry that I will use here is a more symmetric version of the split-bottom experiment, 
which is depicted in Fig. \ref{f1}. The bottom of the container defines the $y$-$z$ plane, which is split along the $y=0$ line. The $x$ direction is perpendicular to the bottom plane. The system is infinite in the $z$ direction (in the three dimensional numerical implementation, periodic boundary conditions  along $z$ will be used). The configuration is symmetric upon a change of sign of the $y$ coordinate, and thus
the center of the shear band is always located at $y=0$. 
Note that this applies to the {\em time averaged} position of the shear band, and not necessarily to an instantaneous configuration. We will come back to this fact in the discussions. This geometry
has been analyzed numerically using frictional, spherical particles in \cite{ries1}. 

To predict, or justify, the form of the shear band we have to make assumptions about the structural response of the
medium. One of the simplest starting points to model a plastic material is the assumption of `perfect plasticity'.
Under this hypothesis, each piece of material behaves elastically at small deformation, i.e., the stress is proportional to the strain, but if some threshold deformation is over passed, the stress remains constant.
It can be shown that if this simple prescription is followed, a singular shear band is obtained that goes from the split line at the bottom 
to the free surface at the top.\cite{unger1} If the threshold deformation is uniform in the sample the shear band is a vertical plane above the 
split line, namely the $y$-$z$ plane. However, if different thresholds are assigned to different points of the sample the shear band obtained is the homogeneous projection in the $z$ direction of a wavy 
line in the $x$-$y$ that goes between the split line at the bottom, and the free top surface. 
This shear surface can be obtained as the one that minimizes the energy generated as `heat of friction' during sliding. 

In Ref. \cite{torok2}, Torok {\em et al.} put forward the idea that this minimal path is not persistent 
in time, since upon some rearrangement of the particles, the thresholds near the shearing surface are likely to change, and the path can move around slightly. In this way, 
a fluctuating path connecting the split point and free surface is obtained. It was thus suggested that the average configuration of this fluctuating path produces the continuum shear band observed in the experiments. 
Simulations based on the two-dimensional version of this idea (in which the $z$ coordinate is eliminated, and the problem is solved in the $x$-$y$ plane, assuming homogeneity along $z$) find a shear band that, after time averaging, is an object that widens from bottom to top. 

An interesting magnitude that has been studied with some detail in this context is the scaling of the width $w$ of the
shear band at the surface as a function of the cell height $H$. In the first reported experimental work \cite{fenis1}
it was found that there is a power law dependence for this quantity, with the shear band width $w$ scaling as $H^{\alpha}$, with $\alpha$  between 0.5 and 1.0.  In the numerical work of Torok {\em et al.} \cite{torok2}, a value close 
to 2/3 was found for $\alpha$ , and it was noticed that this is the analytical value known to occur in the case of directed polymers (DPs) in $1+1$ dimensions.\cite{dirpol} After this, it seems that people have taken the value of 2/3 as a reference, and systematically try to scale the results of experiments or simulations with this value. 

In my view there is not any strong reason why this should be the exact scaling exponent. 
The only point to be remarked is that the problem of DPs at zero temperature \cite{dirpol} is in essence
very similar to the case of the fluctuating shear line in two dimensions. The main difference is 
that in the case of polymers the whole distribution of thresholds (and not only close to the shear line) is
updated in every time step. An additional difference is that the model in \cite{torok2} is not `directed'. However
it is known that for polymers this restriction is not crucial either. The fluctuation of the tip of DPs 
at the surface of the system is analytically known to grow as $H^{2/3}$. However, the 
problem of the shear band is a continuous three dimensional one, and this makes the 
comparison with the DP rather obscure.

Much less experimental information exists on the width of the shear band inside the material, namely $w(x)$, for $0<x<H$. This is in part originated in the difficulty to observe the movement of the particles in the bulk, although some advances have been made in this direction.\cite{sakaie1} 
In Ref. \cite{ries1}, Ries {\em et al.} did numerical simulations using a rather realistic three dimensional model of spherical
particles interacting through friction contact forces. I consider that the results in this work complement nicely
the available experimental measurements. 
In these simulations, 
in addition to the (approximate) 2/3 power scaling of $w(H)$, a quarter of a circle profile was adjusted for the width in the interior of the material, namely $w(x)\sim \sqrt{x(2H-x)}$.

It would be interesting to compare these results with the equivalent ones obtained with the previously mentioned model of fluctuating shear lines \cite{torok2}. Unfortunately, the results in Ref. \cite{torok2} are not very detailed concerning the width across the whole sample. 
Coming to the problem of DPs, to my surprise I have not found a detailed study of the full profile either, and then I set up a simulation code to study this case too. The results obtained for DPs will be presented below. 

The DP model, or the model of fluctuating shear line of Ref. \cite{torok2} consider that instantaneously, the shear band connects the split-bottom point and the free surface by a single line that fluctuates in time. Experimentally, there is no evidence that this is the case. In fact, even for the lowest velocities that can be measured, the velocity profiles that are observed seem to be continuous. This is one reason to try some model which is a bit more realistic and considers the possibility of continuous deformations, both in time and space.
Another reason to search for a more realistic model is the fact that
the DP model, and the model of fluctuating shear line of Ref. \cite{torok2} are eminently two dimensional, and cannot be easily generalized to three dimensions.\cite{3dpol} In fact, in three dimensions, an instantaneous slip surface cannot occur throughout the system, unless its profile is independent of the third ($z$) dimension. 
The problem of the width of a shear band we are discussing is really a three dimensional one. Although it can be 
investigated by two dimensional models (defined in the $x$-$y$ plane in Fig. \ref{f1}), it is not clear 
if there are crucial issues associated to the third direction. In this sense, note that the problem is homogeneous in the $z$ direction only on average, but not instantaneously due to the randomness of the material.
The model presented in the next Section can be implemented in three dimensions, and I will present results both for two and three dimensions.

\section {A Mesoscopic Scale Model}

The model studied here is a simplification over the model introduced in Ref. \cite{jagla} to study
the finite width of shear bands in the presence of structural relaxation. Here this structural relaxation will be absent.
I consider a two dimensional planar geometry in  `mode III' configuration, in which an
out-of-plane displacement $u(x,y)$ is defined. The strain field is defined by the two-component vector 
$e_1=\partial u/\partial x$, $e_2=\partial u/\partial y$.

A perfectly elastic two dimensional material is modeled through the local free energy 
$f=B(e_x^2+e_y^2)$, where $B$ is related to an elastic constant of the material. 
A plastic material possesses a plastic strain field
$e_i^{pl}$, ($i=x$, $y$), in such a way that the local free energy reads

\begin{eqnarray}
f(x,y)&=&B\left [(e_x-e_x^{pl})^2+(e_y-e_y^{pl})^2\right]\nonumber\\
F&=&\int f(x,y) dxdy
\label{f}
\end{eqnarray}
The time evolution of $u(x,y)$ can be considered to be determined in the quasi-static limit by
a relaxational first order equation of the form

\begin{equation}
\frac{du(x,y)}{dt}=-\lambda\frac{\delta F}{\delta u(x,y)}
\label{edet}
\end{equation}

The model must be complemented by the time evolution of the plastic strain.
In principle, a totally
deterministic model for the evolution of $e_i^{pl}$ could be tried. For instance, we can postulate
that if the local energy exceeds some threshold value, the system adjusts its plastic strain to reduce
the local free energy. However, all attempts to implement
this kind of deterministic model led in the present geometry to a singular slip surface, i.e., to a shear band of zero width.
In order to obtain a shear band of finite width, the totally averaged, deterministic evolution
does not seem to be enough, a stochastic component must remain.\cite{fp} There are different possibilities for
the  choice of the stochastic evolution, but I think the kind of results that are obtained is rather
independent of the details. The model adopted here is the following. I start with a zero local plastic 
strain. Upon evolution according to Eq. \ref{edet}, the strain field $e_i$ changes, and also does
the local energy $f$ (Eq. \ref{f}). The plastic strain is fixed to zero until $f$ overpasses some local 
threshold value $f_{th}(x,y)$, then the value of 
$e^{pl}(x,y)$ is set instantaneously to the local value of the strain. In this way the local free energy is
relaxed to zero. In addition, a new, random value of $f_{th}(x,y)$ is chosen. This accommodation of the plastic
strain can be considered to be the plastic events that occur in the system, which allow to produce a potentially very
large deformation, at the cost of a moderate energy consumption.

\section{Results}

\subsection{Directed Polymers}

\begin{figure}[h]
\includegraphics[width=8cm,clip=true]{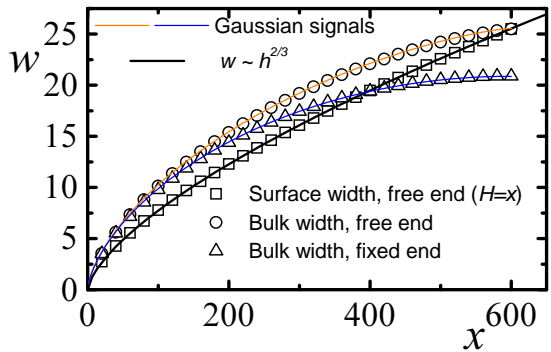}
\caption{(Color online) Squares: Numerical results for the average fluctuation of the free end of a directed
polymer as a function of the number of its elemental
pieces. The analytical 2/3 power (thin continuous) is very well reproduced once the number of elemental pieces becomes large. The full profile of a free end polymer (circles), and the half profile
of a fixed end polymer (triangles) with $H=600$ are also plotted (for better visualization not all points are plotted). Continuous lines: 
fitting using the random Gaussian signals of Ref. \cite{rosso} (Eq. \ref{w3}).
}
\label{f2}
\end{figure}

First of all I will present the results obtained using the DP model.
As a first check, I reproduced with high precision the
analytical 2/3 power for the scaling of the width at the surface 
if the number of elements of the polymer
is sufficiently large, as can be seen in Fig. \ref{f2}.
The full curve $w(x)$ of the largest polymer studied is also plotted in that figure. It coincides at the surface with
the previous curve, by definition, but is systematically different in the bulk.
To my knowledge, there is no analytical expression for the $w(x)$ profile of a DP. 
Note that the half-profile for a DP fixed at both ends (plotted also in Fig. \ref{f2}) shows clear differences with the free end case: the width of the free end case is systematically larger than the fixed end case, they only coincide asymptotically close to the fixed point. Note also that the curve for the two fixed ends case 
has clearly zero slope at the middle point, which is obviously due to symmetry reasons, but this is not obviously the case
for the free end case. 

To analyze these results in more detail, it turned out to be very successful to apply 
the theory of Gaussian signals developed in Ref. \cite{rosso}. This consists in the study of the properties of a
self-affine stochastic process, submitted to the appropriate boundary conditions, assuming 
a decomposition into uncorrelated Fourier modes of Gaussian amplitude. Although the hypothesis of the modes being uncorrelated is usually not satisfied, this analysis is known in particular cases to provide results
that fit the exact data with high accuracy.\cite{error}
The simplest case to start with is that of the fixed ended polymer. In this case it is assumed that the 
stochastic process can be decomposed in Fourier modes in the form:

\begin{equation}
Y(x)=\sum_{n=1}^\infty g_n A_n \sin (n\pi x/H)
\end{equation}
where $Y(x)$ is the $y$ coordinate of the polymer at height $x$ (see Fig. 1), $g_n$ are uncorrelated random numbers
with zero mean and unitary variance, $H$ is the distance between the two fixed ends of the polymer, and $A_n$ is the 
amplitude of the corresponding mode. For self affine processes with scaling properties, the amplitudes $A_n$ must 
scale as a power of $n$, namely $A_n=C_0 n^{-\gamma}$, where an $H$-dependent global constant was introduced. By multiplying and taking averages over realizations we get 
the statistical width $w$ at distance $x$ of one fixed end as
\begin{equation}
w^2(x)=C_0^2\left( \sum_{n=1}^\infty n^{-2\gamma} \sin^2 \left(n\pi x/H\right) \right)
\label{w2}
\end{equation}
where I have made use of the fact that different $g_n$ are uncorrelated.
The exponent $\gamma$ can be related to our exponent $\alpha$ for the expected scaling of the stochastic process at the
surface. It is not difficult to show that $\alpha=\gamma-1/2$. Thus we can write 

\begin{equation}
w^2(x)=C_0^2\left( \sum_{n=1}^\infty n^{-2\alpha-1} \sin^2 \left(n\pi x/H\right ) \right)
\label{w22}
\end{equation}
 
As an example, note that the thermally fluctuating elastic line problem, has $\alpha=1/2$, the sum in Eq. (\ref{w22}) gives

\begin{equation}
w(x)=\frac{\pi C_0}{\sqrt 2 H}\sqrt{x(H-x)}
\end{equation}
i.e., the well known half circle, which is the exact result in this case. 
For DPs we have $\alpha=2/3$, and the result for $w(x)$, fitting the global factor $C_0$ is shown on top of the numerical results in Fig. \ref{f2} and cannot be distinguished from the numerical data. 

Going to the case of the free end polymer, note that in the problem of thermally fluctuating elastic line, only modes
that reach the surface with zero derivative must be considered. This leads to a profile of the form
\begin{equation}
w^2(x)=C_0^2 \left (\sum_{n=1,3,5,...}^\infty n^{-2\alpha-1} \sin^2 \left(n\pi x/2H\right)\right)
\label{w2p}
\end{equation}
where now the sum is over half the modes we had before. The result obtained scales as
\begin{equation}
w(x)\sim\sqrt{x}
\end{equation}
In particular, this profile does not have zero derivative at the surface, although it is constructed with modes that have 
zero derivative individually. 

Coming back now to the free end DP polymer, 
an attempt can be made to fit the curve of an open end polymer weighting appropriately the modes with zero value at the surface, and those with zero derivative, keeping the self-affine nature of both of them individually. This corresponds to 
a width function of the form:

\begin{equation}
w^2(x)=C_0^2\left(\sum_{n}^\infty (1+\beta (-1)^n)n^{-2\alpha-1} \sin^2 (n\pi x/H)\right)
\label{w3}
\end{equation}

The best fitting to the numerical results is obtained when we choose $\beta=1/2$, and $C_0$ as the same value used to fit the fixed end case.
The fitting, shown in Fig. \ref{f2}, cannot be distinguished from the numerical results either. Note that having the same value of $C_0$ means that the asymptotic form of $w(x)$ for $x\to 0$ is exactly the same for the free end and open end 
situations (this is because the alternating in sign $\beta$ term in (\ref{w3}) introduces for small $x$ a contribution that is a larger power of $x$ than the main term). 

\begin{figure}[h]
\includegraphics[width=8cm,clip=true]{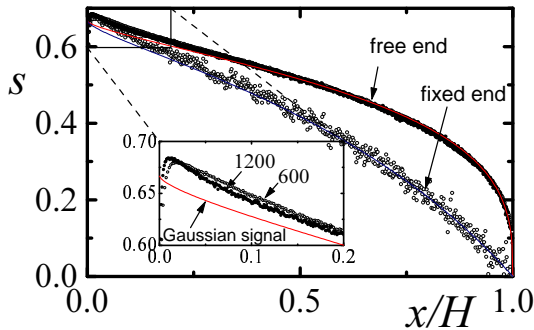}
\caption{(Color online) Logarithmic derivative of 
the bulk width data of free end DPs with $H=600$ and 1200, and fixed end DP with $H=600$, and the 
corresponding fitting according to random Gaussian signals (\cite{rosso}). Note that for free end DPs, the derivative vanishes at the surface at $(H-x)^{1/3}$. Note also that the data for small arguments tend to approach the trend of the continuous line as size is increased, although very slowly (the blow-up shows only data for the open end case).
}
\label{f3}
\end{figure}

A very sensitive quantity that can be used to evaluate the results just presented is the
local slope $s$ of the $w(x)$ curve, defined as $s(x)=d \ln w(x)/ d\ln x$. These results, both
from numerics and using the fitting with Gaussian signals are shown in Fig. \ref{f3}. 
Two facts must be noticed:
first, for open end polymers, the slope vanishes at the surface as $(H-x)^{1/3}$, more abruptly than the linear vanishing 
of the fixed end case at the mid point. Second, although the convergence is slow with system size, data seem to indicate
that the form of $w(x)$ for $x\to 0$ is $w(x)\sim x^{2/3}$, i.e., it has the same exponent that the surface scaling (which is, by the way, a property of the Gaussian signal too). I 
will come back to the meaning of this fact in the last section.

At present, I do not have any explanation that justifies the accuracy of the fitting proposed, and neither an explanation for the remarkably  value of $\beta$ obtained in the fitting of the open end polymer.
The results just presented for DPs will serve as a reference for comparing those obtained with the mesoscopic plastic model.

\subsection{Plastic mesoscopic model: two dimensions}

In a first stage, I implemented the plastic mesoscopic model on a two dimensional rectangular 
geometry corresponding to the $x$-$y$ plane in Fig. \ref{f1}, in which the border of the right half is given some velocity $v$, and the border of the left half is given the opposite velocity $-v$. 
The value of $v$ is set small enough for the total velocity field to be simply proportional to this value.

For the split bottom geometry, the top surface is left free, i.e., a condition of zero normal stress is applied.
In some results presented, the top surface will be also split in the same manner that the bottom surface, and this will be a `split top-bottom' configuration.
Note that although I use the expressions `top' and `bottom' to agree with the experimental device, in the present case
I do not include a gravitational field that singles out a vertical direction. This choice is made as the numerical simulations in \cite{ries1} indicate that gravity plays only a minor role in the structure of the shear band. 
Simulations were performed for different values of the total height of the system $H$, and in each case the size along the $y$ direction was chosen sufficiently large for the shear band not to be affected by the lateral walls (typically,
the size along $y$ is around ten times the expected value of $w$ at the surface).

Starting from rest, with both $e_i$ and $e_i^{pl}$ set to zero, the simulation is initiated by applying a finite 
velocity $\pm v$ to the left and right borders. At the first stages of the simulation, the system is seen to respond elastically, since in no place the value of $f_{th}$
is over passed. At some moment, $f_{th}$ is over passed for the first time at some place, the corresponding value of 
$e_i^{pl}$ changes, and this feeds back to modify the evolution of $e_i$.
After many plastic events at different spatial positions, the system reaches a stationary state, 
(on average, since stochastic fluctuations of the plastic strain remain), and at this stage statistics can be accumulated to get the average velocity field. 
In the simulations presented, the threshold values $f_{th}$ are chosen randomly in the interval
$0.5 f_{th}^0<f_{th}<1.5 f_{th}^0$ where $f_{th}^0$ is a reference value. Dimensionless units are set by taking
$f_{th}^0\equiv 1$ and $\lambda B\equiv 1$ in Equations (\ref{f})-(\ref{edet}). In these units, the small velocity used
to drive the system is $10^{-4}$, and the time step y 0.1.

\begin{figure}[h]
\includegraphics[width=8cm,clip=true]{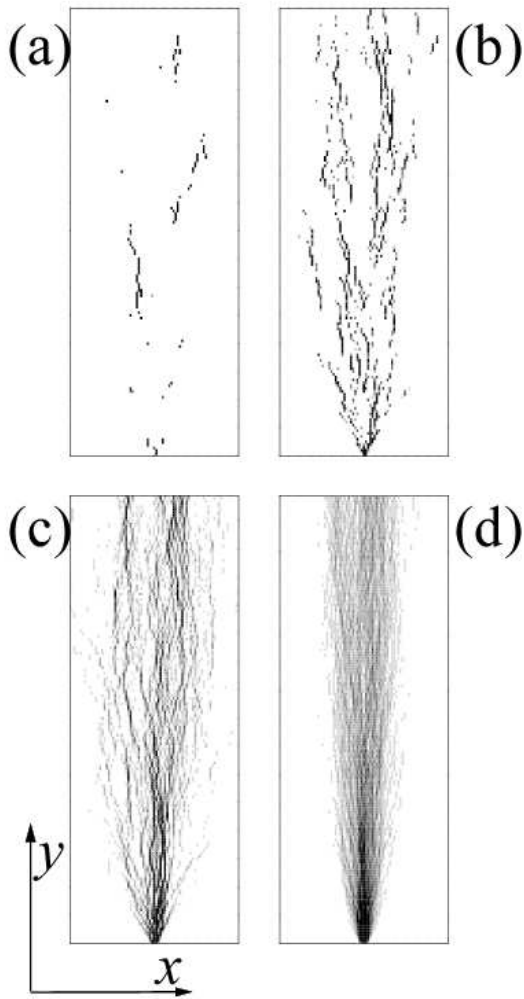}
\caption{Schematics of plastic deformation in the horizontal direction ($\partial u/\partial x$) accumulated over progressively larger 
periods of time $\Delta t$, namely $\Delta t/t_0=10^2$, $10^3$, $10^4$, and  $1.5\times 10^5$, from (a) to (d). 
Darker regions indicate zones with larger accumulated plastid deformation. Note that the plastic events appear at rather isolated positions (a), 
that then mersge to generate pieces of shear lines (b-c). Over very large periods of time
the fluctuating behaviors averages out, and a stationary, continuous shear band is observed (d).
}
\label{f4}
\end{figure}

The shear band forms as an effect of the sequence of plastic events in which the values of $e_i^{pl}$
are adapted at different points in space. These events are correlated in space, in general. A plastic slip
occurring at some position $(x,y)$ reinforces the possibility of a similar event occurring above 
or below, at the same $y$ coordinate, and displaced in the $x$ direction. However, the stochastic nature of the model makes this overall tendency not to be strict.
Accumulated plastic deformations over progressively larger periods of time, 
shown in Fig. \ref{f4}, allow to understand the mechanics of the process. 
On very short time scales (Fig. \ref{f4}(a)), individual events at isolated positions are observed. At intermediate time scales (Fig. \ref{f4}(b) and (c)), we see how these events are correlated in space to generate a sort of finite length shear lines. Note however, that contrary to the fluctuating shear line model, here the plastic deformation 
does not accumulate at a single shear line, irrespective of the time interval considered.
For the largest time scales (Fig. \ref{f4}(d)) we can see the tendency for the the fluctuating behavior
to be averaged out, and for sufficiently large time intervals we get eventually a well defined, continuous shear band. 

\begin{figure}[h]
\includegraphics[width=8cm,clip=true]{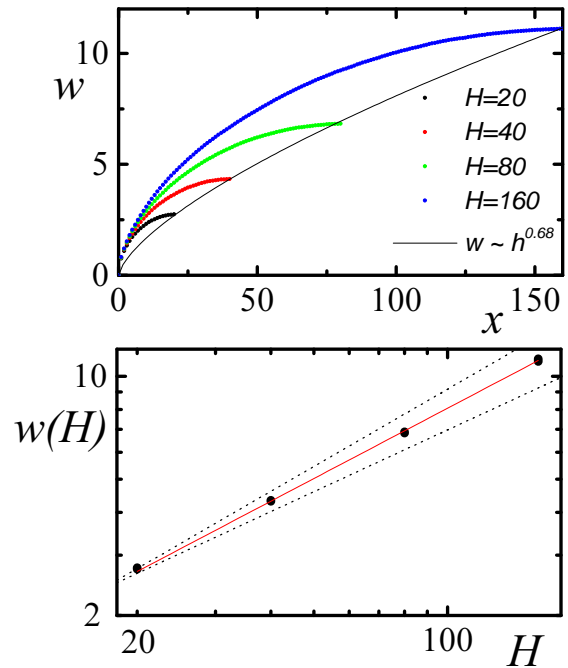}
\caption{(Color online) (a)Full profile of the width of the shear band across the sample for samples of different size. 
(b)Evolution of the value of $w$ at the surface with system size. The error bar is smaller than symbol size. The continuous line is the best fitting, with an 
exponent 0.68. Dotted lines have slope 0.60 and 0.75, for comparison.
}
\label{f51}
\end{figure}

\begin{figure}[h]
\includegraphics[width=8cm,clip=true]{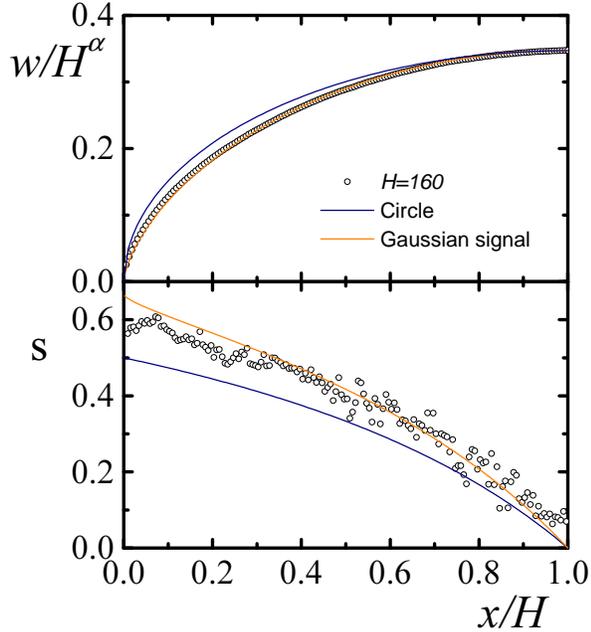}
\caption{(Color online) 
(a)The results of the previous figure scaled using an exponent $\alpha=0.68$. The fitting using Gaussian signals and a quarter of circle profile are shown. (b)Logarithmic derivative of the data in (a), and the corresponding curve from
Gaussian signals and for a circle.
}
\label{f52}
\end{figure}

Now I will focus on the properties of the averaged shear band in this stationary situation. 
In Fig. \ref{f51}(a) I plot the full profile for the width $w$ of the shear band obtained in systems of 
different heights $H$.
These data can be very well scaled if plotted as $w/H^{\alpha}$
against $x/H$. This is basically the same result obtained in experiments and in other simulations, extended here to the full profile inside the sample. The best fitting for the power $\alpha$ gives $\alpha=0.68\pm 0.02$ (see Fig.
 \ref{f51}(b)).
Applying to this case the fitting using Gaussian signals, we get the fitting shown in Fig. \ref{f52}. This is 
obtained using expression (\ref{w3}) with $\alpha=0.68$ (taken from the surface scaling), and $\beta=0$.
The corresponding logarithmic derivative of the data and the fitting are shown also in Fig. \ref{f52}(b). Although a slightly different from zero value of $\beta$ gives a somewhat better fitting, I think it is within the numerical uncertainty. 
The plot of the logarithmic derivative of the data (Fig. \ref{f52}(b)) shows that the form of the 
curves close to the fixed point may correspond, in the large system size limit, to have the same exponent as the one of the surface scaling.

\begin{figure}[h]
\includegraphics[width=8cm,clip=true]{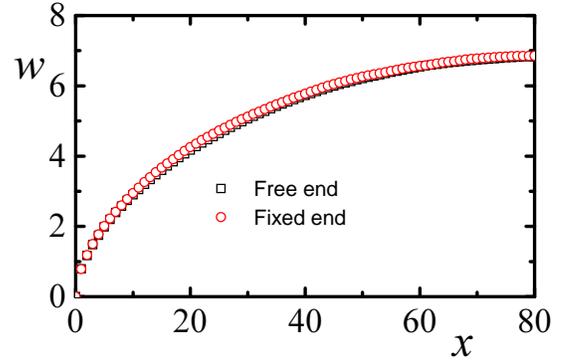}
\caption{(Color online) Comparison between the profile obtained for a split bottom configuration (free end case), and a split 
top-bottom configuration (fixed end case), for $H=80$. No systematic differences can be observed 
within the numerical errors.
}
\label{f7}
\end{figure}

Note that the $\beta\sim 0$ value for the free surface case means that probably in the present model there are not significant difference between the split bottom and split top bottom geometry of double size. To confirm this fact, 
in Fig. \ref{f7} I show the results for the split bottom, and split top-bottom configuration with $H=80$, where it can be seen than
there are not any significant differences between the two.
This results was not obvious from the beginning, since the symmetry of the split top bottom geometry with respect to the horizontal middle plane exists only on average, and not instantaneously.
The present finding agrees with the observation in Ref. \cite{ries1} of a profile that hits the surface with zero derivative (in that work however, it was not checked that the full profile was identical to half profile of a
split top bottom case with double size). 
In Ref.\cite{ries1} it was also observed that the full profile of the $w(x)$ could be nicely fitted with a
quarter of a circle profile. An attempt to do this in the present case shows (see dashed line in Fig. \ref{f52}) 
apparent differences between this
function and the numerical data that seem to be much larger than those reported in \cite{ries1}. I will come to this attempt of fitting in the next section.

In addition to the plot of the width of the shear band, it is interesting to evaluate the full profile of the velocity at some typical vertical position, i.e, to consider the velocity $v(y)$ at some fixed value of $x$. 
Previous numerical and experimental results show that to a good approximation this profile can be fitted by an
error function. The experimental data do not have sufficient precision so as to observe systematic differences with respect to a perfect error function. On the other hand, the numerical simulations in \cite{ries1} 
show profiles that go above
the error function for horizontal positions far away of the center of the shear band.
It is not totally clear however if this difference will persist in infinitely long runs, as the difference slightly reduces as the simulation time increases.

\begin{figure}[h]
\includegraphics[width=8cm,clip=true]{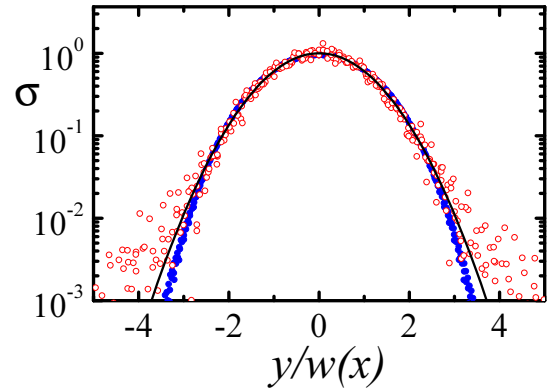}
\caption{(Color online) Derivative of the velocity profile $\sigma(y)\equiv dv(y)/dy$
for the directed polymer (full symbols) and the plastic mesoscopic model (open symbols).
Data at $x=H/4$, $x=H/2$, $x=3H/4$, an $x=H$ are put all together by scaling the $y$ axis with the corresponding 
values of the width $w$. The deviation of DPs from a Gaussian (in defect for large values of $x$)
is clearly visible. The quality of the data for the plastic model is not sufficient to make a statement about 
deviations with respect to a Gaussian.
}
\label{f8}
\end{figure}

Given the velocity profile $v(y)$ at some fixed $x$ position, its $y$ derivative will show a Gaussian form. This quantity, noted $\sigma (y)$ is presented in Fig. \ref{f8}
for the plastic model, and for DPs.  Data at the surface of the system, and in some
particular positions in the bulk are included to show that there are no systematic differences between surface and bulk. The $y$ axis is scaled in each case by the corresponding value
of $w(x)$, and the normalization $\sigma(0)=1$ is used,  in such a way that the best Gaussian fitting is given in all cases by the function $\exp(-x^2/2)$.
The results for DPs fall clearly below the Gaussian fitting for sufficiently large values of the argument, namely in the tail of the distributions. This is a well known result.
Results for the full model indicate a very good Gaussian fitting for $|y|\lesssim 2 w(x)$.
Data seem to indicate a slight deviation in excess for larger values of $y$. However, these result require more massive simulations to be confirmed.

\subsection{Plastic mesoscopic model: three dimensions}

The results of the previous section indicate that the two dimensional plastic mesoscopic model has similarities and differences with the fluctuating shear line, or the DP model, and with results of other kind of simulations \cite{ries1}. In particular, the result obtained for the exponent 
$\alpha =0.68 \pm 0.02$ is remarkably consistent with the analytical $2/3$ exponent of DPs. Experimentally, the obtained exponent is consistent with a 2/3 value, but to my knowledge an independent determination has not been attempted, and 
probably values within $\pm 0.1$ of 2/3 will be equally consistent with the experiments.

However, an important point of discrepancy between the present results and atomistic numerical simulations in Ref.
\cite{ries1} is the form of a quarter of a circle profile found for $w(x)$ in \cite{ries1}. The results of the previous Section (see Fig. \ref{f52}) seem to depart from this behaviors in a larger extent than the results in \cite{ries1}.

One of the possible reasons for this discrepancy is the fact that the results in \cite{ries1} and experiments in general are fully three dimensional, while all the results presented
up to now here are two dimensional. It is important to emphasize that although in the geometry studied
there is translation symmetry {\it on average} along the $z$ direction, the stochastic nature of the problem implies that fluctuations are not invariant along this direction. Actually, the setting up of a model that can be implemented in three dimensions was one of the motivations of the present research. 

A full implementation of the mesoscopic plasticity model in three dimensions requires the use of a three component displacement field $u_i(x,y,z)$, $i=x,y,z$. 
This presents some technical challenges that I have not overcome yet. However, a partial implementation of the three dimensional case is possible, and the results obtained are already very interesting.

\begin{figure}[h]
\includegraphics[width=8cm,clip=true]{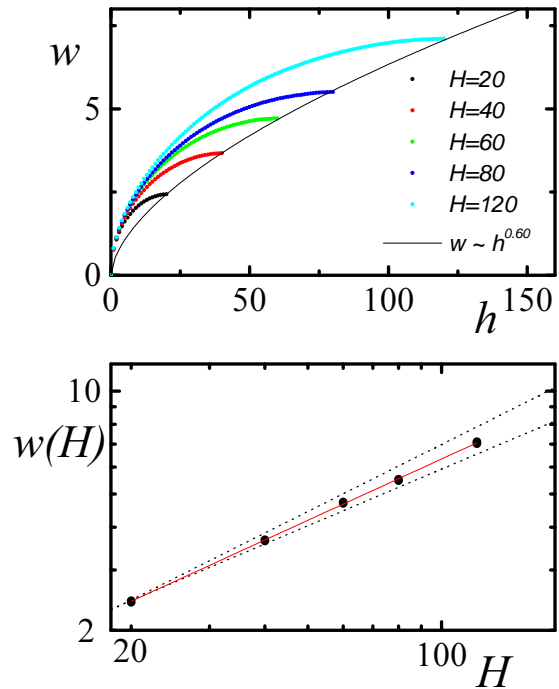}
\caption{(Color online) (a)Three dimensional data for different values of $H$.
(b)Evolution of the value of $w$ at the surface with system size. The error bar is smaller than symbol size. The continuous line is the best fitting, with an 
exponent 0.60. Dotted lines have slope 0.55 and 0.65, for comparison.
}
\label{f3d1}
\end{figure}

\begin{figure}[h]
\includegraphics[width=8cm,clip=true]{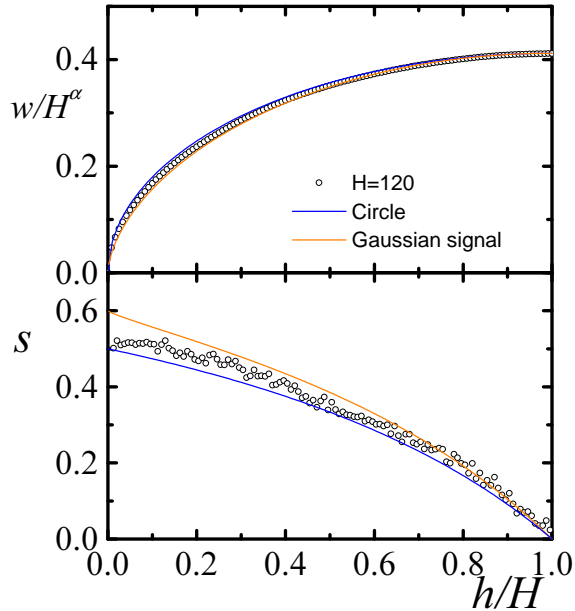}
\caption{(Color online) (a)The data of the previous figure corresponding to the largest system size ($H=120$) are shown 
together with a fitting using a Gaussian signals with exponent 0.60  and a quarter of circle profile. (b) Logarithmic derivative of the data in (a), and the corresponding curves from
Gaussian signals and for a circle.
}
\label{f3d2}
\end{figure}

The idea is to keep working with a single component displacement field $u$ which is the displacement along $z$ direction, keeping $u_x=u_y=0$. The model can thus be described as a collection of planes in which a set of equivalent two dimensional problems are defined, and where the plastic thresholds are chosen independently for each plane. 
The fundamental variable is then some $u(x,y)_i$ where index $i$ labels the plane along the $z$ direction.
Planes are connected to nearest neighbor ones by purely elastic forces, in such a way that the total energy of the system is now

\begin{equation}
F=\sum_i \left (F^i +\frac{k_z}{2} \int dxdy (u_i-u_{i+1})^2\right )
\end{equation}
where $F^i$ is the previous two dimensional free energy for plane $i$.
Periodic boundary condition are applied along $z$. The size of the system along $z$ should be chosen large enough.
To keep a consistent implementation, this size is always chosen to be equal to the total thickness $H$ of the system. The coupling constant $k_z$ along $z$ is chosen to be equal to the linear elastic constant within the plane. Note that no plastic threshold is implemented for the springs along $z$. 
This version of the model was simulated in exactly the same manner as the two dimensional case. The results for the
width function $w(x)$ for systems of different sizes are shown in Figs. \ref{f3d1} and \ref{f3d2}. Data for different values of $H$ can again be scaled by plotting $w/H^{\alpha}$ vs $w/H$. The fitting of $\alpha$ gives $\alpha=0.60\pm 0.02$.

Two important differences appear between the three dimensional and the previous two dimensional simulations. First 
of all, the scaling of data for systems of different thickness is obtained using an exponent smaller than that of the two dimensional case, namely $\alpha^{3D}\sim 0.60$, $\alpha^{2D}\sim 0.68$. The precision of the present simulations
allows to claim that this difference is significant.
In my view other numerical simulations as those in \cite{fenis1} or experiments have not reached yet the point to tell
the growing exponent with a precision of $\pm$ 0.1. I think a value close to 0.6 is consistent with the precision of
accessible experimental results. More experimental work is needed concerning this point. The second difference between two and three dimensions is the overall shape of the $w(x)$ function. The form of this function in 3D is much closer to a quarter of circle profile (see Fig. \ref{f3d2}). 
This may be an interesting observation when comparing with the results in \cite{ries1}, for instance.
However, I have to mention that the fitting to this functional form does not seem to improve as the system size is increased, systematic differences apparently exist. 

Related to this behavior, there is an interesting point to emphasize. Given a set of curves $w(x)$ for different thicknesses $H$, if the scaling hypothesis holds we have 
\begin{equation}
w(x)=w_0(y/H)H^{\alpha}
\end{equation}
where $w_0$ is an $H$-independent function. If we assume $w_0$ behaves as a power law (with exponent $\alpha '$)
for small values of its argument, we have
\begin{equation}
w(x\to 0)\sim (x/H)^{\alpha '}H^{\alpha}=x^{\alpha '}H^{\alpha-\alpha '}
\end{equation}
Note that unless $\alpha=\alpha '$, the behavior of $w$ close to the origin will depend on the value of $H$, no matter how large this is. In particular, $w(x\to 0)$ will increase without limit (if $\alpha>\alpha '$) or decrease to zero (if $\alpha<\alpha '$) as $H$ increases. 
This would be in my view a very strange behavior, although I have no rigorous argument against it. The DP curves shown in Figs. \ref{f2} and \ref{f3} are in fact compatible with a $2/3$ power close to the fixed point, and for the two dimensional plastic mesoscopic model a value of $\alpha'=\alpha$ seems to be compatible with data in Figs. \ref{f51} and \ref{f52}. The situation is less clear in the three dimensional case. The value of $\alpha$ was found to be around 0.60. However, the value of $\alpha' $ from Fig. \ref{f3d2} seems
to be consistently closer to 0.5 (which, by the way, is why the quarter of circle profile is accurate). It is not clear if the two exponents will show a tendency to become closer in larger system size simulations of if they will remain different.

\section {Summary and Conclusions}

In this paper I presented results obtained for the `split-bottom' geometry using a mesoscopic model for the plasticity of an amorphous material. Two and three dimensional results were reported.  
In particular, the exponent for the width $w$ of the shear band at the surface of the system for two dimensions was found to be 0.68 $\pm$ 0.02, comparable with the analytical 2/3
exponent obtained for directed polymers. In three dimensions the exponent was found to be somewhat slower, namely 0.60 
$\pm$ 0.02. I want to recall however that the implementation of the three dimensional case was only partial, and a full 
modeling would be necessary to verify if this value remains valid in the full implementation. In two dimensions, the full profile of the width in bulk $w(x)$ was found to be well described by a random Gaussian signal, and in particular it was found that
no systematic differences appear between a system with a free top surface and system with a split surface at the top. This was not an obvious point from the beginning as the reference system for this study, namely the directed polymer, shows systematic differences between the two cases that I have addressed and described in terms
of the theory of random Gaussian signals. In three dimensions, it was observed that the form of $w(x)$ can be fitted
reasonably well by a quarter of circle profile, a result that has been obtained in previous work in a related model 
\cite{ries1}. A caveat about whether this scaling can continue to be valid for arbitrarily large systems was put forward.

The number of experimental results that the present model describes, added to the very simple ingredients on which
it is constructed,
places it as a good candidate in which the deep origin of the regularities emerging from experiments can be elucidated. One promising
route would be trying to generate continuous differential equations that describe the evolution of probability densities
of the stochastic model. This is an interesting prospect for future work.
In addition, I think that the results presented can be a motivation for experimentalists to obtain more accurate data
of the quantities at play, to investigate how the values presented here fit the experiments.

\section{Acknowledgments} 

Fruitful discussions with A. Kolton, A. Rosso, M. Falk, W. Losert and T. Unger are greatly acknowledged.
This research was financially supported by Consejo Nacional de Investigaciones Cient\'{\i}ficas y T\'ecnicas (CONICET), Argentina. Partial support from
grants PIP/5596 (CONICET) and PICT 32859/2005 (ANPCyT, Argentina) is also acknowledged.



\begin{thebibliography}{15}
\bibitem{3dpol}Although the problem of directed polymers can be easily generalized to 2+1 dimensions, its 
relevance to the shear band problem is much less obvious in this case.
\bibitem{fp}Once the stochastic model is defined, one can try to obtain from it a smooth differential equation that describes the evolution of probability density in the stochastic model. This is like obtaining the Fokker-Planck equation from
a Langevin dynamical equation. In the present case I have no clues on how to obtain this effective equation. This is a very interesting issue to pursuit in the future.
\end{thebibliography}
\end{document}